\documentclass[fleqn,twoside]{article}
\usepackage[headings]{espcrc2}

\readRCS
$Id: espcrc2.tex,v 1.2 2004/02/24 11:22:11 spepping Exp $
\usepackage{graphicx}
\usepackage[figuresright]{rotating}

\newcommand{\AmS}{{\protect\the\textfont2
  A\kern-.1667em\lower.5ex\hbox{M}\kern-.125emS}}

\title{Muon g-2, rare decays $P \to l^+l^-$ and transition form factors $P \to \gamma\gamma^*$}

\author{A.E. Dorokhov\address[MCSD]{Joint Institute for Nuclear Research,
Bogoliubov Laboratory of Theoretical
Physics, \\  141980 Dubna, Moscow region, Russian Federation; \\
Institute for Theoretical Problems of Microphysics, Moscow State University, \\  RU-119899, Moscow, Russian Federation}%
        \thanks{The author acknowledges partial support from the Scientific School grant 4476.2006.2.}}

\runtitle{Muon g-2, rare decays $P \to l^+l^-$ and transition form factors $P \to \gamma\gamma^*$}
\runauthor{A.E. Dorokhov}

\begin{document}

\begin{abstract}
Measuring the muon anomalous magnetic moment $g-2$ and the rare decays of light pseudoscalar mesons into lepton pair $P\rightarrow l^{+}l^{-} $
serve as important test of the standard model. To reduce the theoretical uncertainty in the standard model predictions the data on the
transition form factors of light pseudoscalar mesons play significant role. Recently new data on behavior of these form factors at large
momentum transfer was supplied by the BABAR collaboration. We comment on the (in)consistency of these data with perturbative QCD expectation. \vspace{1pc}
\end{abstract}

\maketitle


The theoretical study and comparison with experimental results of the muon anomalous magnetic moment $g-2$ (see for review
\cite{Miller:2007kk,Passera:2007fk,Dorokhov:2005ff,Jegerlehner:2009ry,Prades:2009qp}), the rare decays of light pseudoscalar mesons  into lepton
pairs \cite{Dorokhov:2007bd,Dorokhov:2008cd,Dorokhov:2008qn,Dorokhov:2009xs} offers an important low-energy test of the
standard model. The discrepancy of the present standard model prediction of the muon anomalous magnetic moment with its experimental
determination \cite{Bennett:2006fi} is $(24.6\pm 8.0)\cdot 10^{-10}$ ($3.1\sigma$) \cite{Prades:2009qp}. The situation with the rare
decays of light pseudoscalar mesons  into lepton pairs became more pressing after recent KTeV E799-II experiment at FermiLab
\cite{Abouzaid:2007md} in which the pion decay into an electron-positron pair was measured with high accuracy using the $K_{L}\rightarrow3\pi$
process as a source of tagged neutral pions
\begin{equation}
B^{\mathrm{KTeV}}\left(  \pi^{0}\rightarrow e^{+}%
e^{-}\right)  =\left(  7.49\pm0.38\right)  \cdot10^{-8}.\label{KTeV}%
\end{equation}
The standard model prediction gives \cite{Dorokhov:2007bd,Dorokhov:2009xs}
\begin{equation}
B^{\mathrm{Theor}}\left(  \pi^{0}\rightarrow e^{+}e^{-}\right)  =\left(
6.2\pm0.1\right)  \cdot10^{-8},\label{Bth}%
\end{equation}
which is $3.3\sigma$ below the KTeV result (\ref{KTeV}). The other modes are given in Table.

\begin{table*}[htb]
\caption[Results]{Values of the branchings $B\left(  P\rightarrow l^{+}%
l^{-}\right)  $ obtained in our approach and compared with the available
experimental results. }%
\label{table2}
\renewcommand{\arraystretch}{1.0} 
\begin{tabular}[c]{|c|c|c|c|c|c|}\hline
$R_{0}$ & Unitary bound & CLEO bound & CLEO+OPE & \cite{Dorokhov:2009xs} & Experiment\\
&  &  &  &  & \\\hline $R_{0}\left(  \pi^{0}\rightarrow e^{+}e^{-}\right)  \times10^{8}$ & $\geq4.69$ & $\geq5.85\pm0.03$ & $6.23\pm0.12$ &
$6.26$ & $7.49\pm0.38$ \cite{Abouzaid:2007md}\\\hline $R_{0}\left(  \eta\rightarrow\mu^{+}\mu^{-}\right)  \times10^{6}$ & $\geq4.36$ &
$\leq6.23\pm0.12$ & $5.12\pm0.27$ & $4.64$ & $5.8\pm0.8$ \cite{Amsler:2008zzb,Abegg:1994wx}\\\hline $R_{0}\left(  \eta\rightarrow e^{+}e^{-}\right)
\times10^{9}$ & $\geq1.78$ & $\geq4.33\pm0.02$ & $4.60\pm0.09$ & $5.24$ & $\leq2.7\cdot10^{4}$ \cite{Berlowski:2008zz}\\\hline $R_{0}\left(
\eta^{\prime}\rightarrow\mu^{+}\mu^{-}\right)  \times10^{7}$ & $\geq1.35$ & $\leq1.44\pm0.01$ & $1.364\pm0.010$ & $1.30$ & \\\hline $R_{0}\left(
\eta^{\prime}\rightarrow e^{+}e^{-}\right)  \times10^{10}$ & $\geq0.36$ & $\geq1.121\pm0.004$ & $1.178\pm0.014$ & $1.86$ & \\\hline
\end{tabular}
\end{table*}

The main limitation for realistic predictions for these processes comes from the large distance contributions of the strong sector of the
standard model where the perturbative QCD theory does not work. In order to diminish the theoretical uncertainties the usage of experimental data on the pion charge and transition form factors are of crucial importance. The first one measured in $e^+e^- \to \pi^+\pi^-(\gamma)$  by CMD-2 \cite{Akhmetshin:2006bx}, SND \cite{Achasov:2006vp}, KLOE \cite{Aloisio:2004bu}, BABAR \cite{Aubert:2009fg} provides the estimate of the hadron vacuum polarization contribution to muon $g-2$ with accuracy better than $1\%$. The second one measured in $e^+e^- \to e^+e^-P$ for spacelike photons and $e^+e^- \to P\gamma$ for timelike photons by CELLO \cite{Behrend:1990sr}, CLEO \cite{Gronberg:1997fj}, BABAR \cite{Aubert:2006cy,:2009mc} are essential in reducing theoretical uncertainties in the estimates of the hadronic light-by-light process contribution to the muon $g-2$ and the decays $P\to l^+l^-$.

In Figs. 1-3 the data on the $\pi^0$, $\eta$ and $\eta'$ transition form factors from the CELLO, CLEO, BABAR collaborations are presented. The BABAR points at $Q^2=112$ GeV$^2$ \cite{Aubert:2006cy} in Figs. 2 and 3 being for the timelike form factor are drawn assuming that the spacelike and timelike asymptotics are equal.


\begin{figure}[ht]
\hspace*{-10mm}\includegraphics[width=0.6\textwidth]{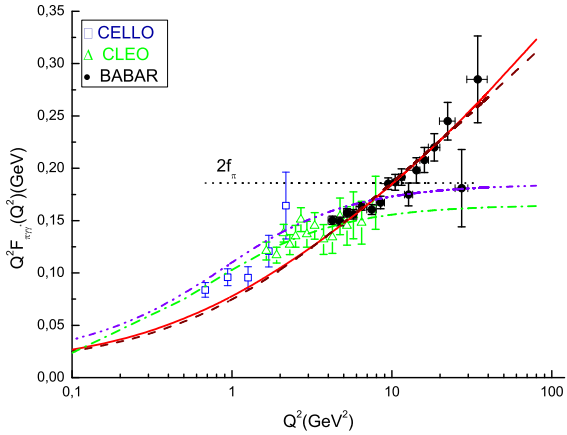}  \vspace*{-10mm}
\caption{{\protect\footnotesize The transition form factor $\protect\pi%
^{0}\rightarrow \protect\gamma^\ast\protect\gamma$. The data are from CELLO
\protect\cite{Behrend:1990sr}, CLEO \protect\cite{Gronberg:1997fj} and BABAR
\protect\cite{:2009mc} Collaborations. The dashed line is massless QCD
asymptotic limit. (The notation for curves is explained in the text.)}}
\label{fig:pi}
\end{figure}

\begin{figure}[ht]
\hspace*{-10mm}\includegraphics[width=0.6\textwidth]{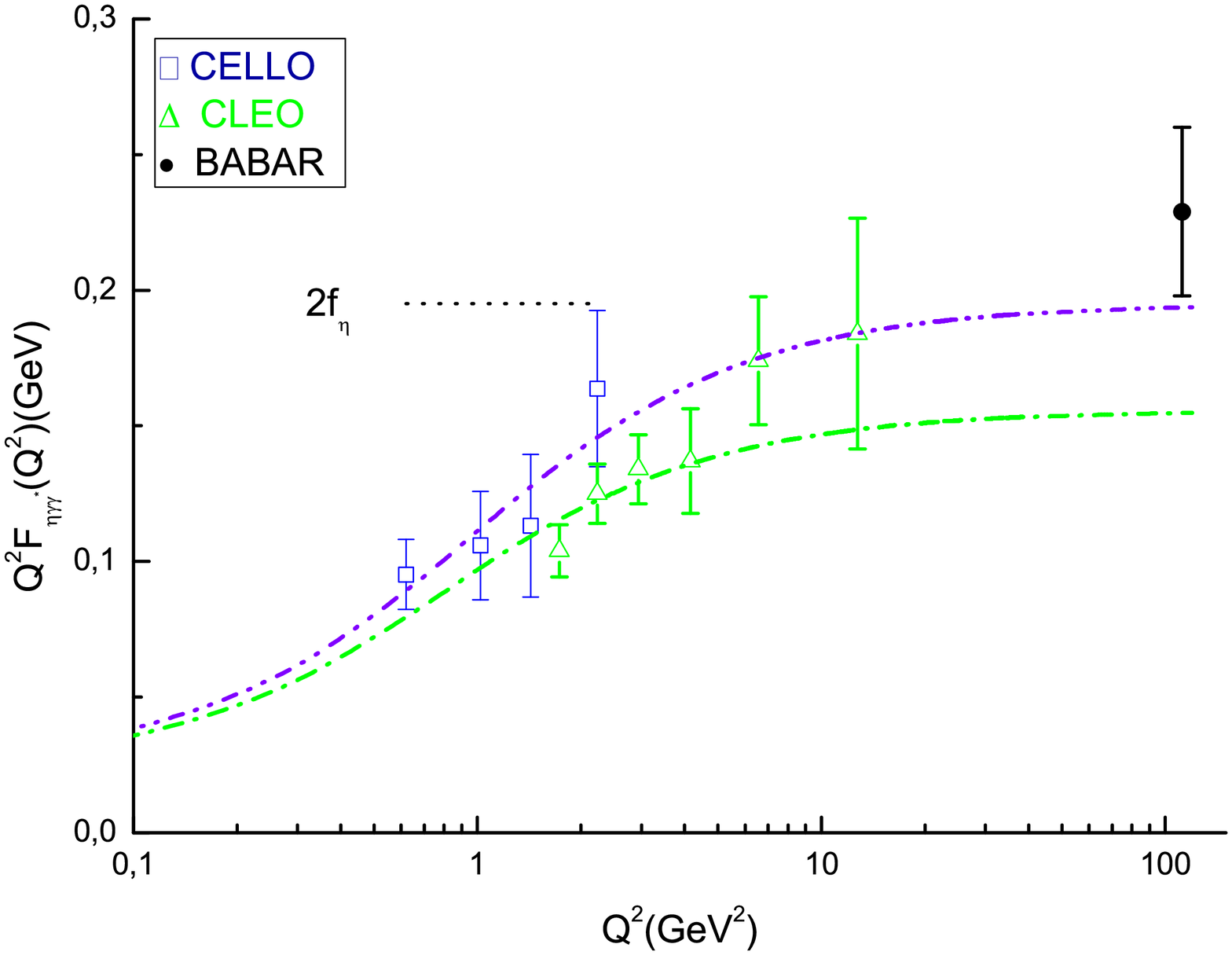}  \vspace*{-10mm}
\caption{{\protect\footnotesize The transition form factor $\protect\eta
\rightarrow \protect\gamma^\ast\protect\gamma$. The data are from CELLO
\protect\cite{Behrend:1990sr}, CLEO \protect\cite{Gronberg:1997fj} and BABAR
\protect\cite{Aubert:2006cy} Collaborations. The
CLEO results obtained in different $\eta$ decay modes are averaged. (The notation for curves is explained in the text.)}}
\label{fig:eta}
\end{figure}

\begin{figure}[ht]
\hspace*{-10mm}\includegraphics[width=0.6\textwidth]{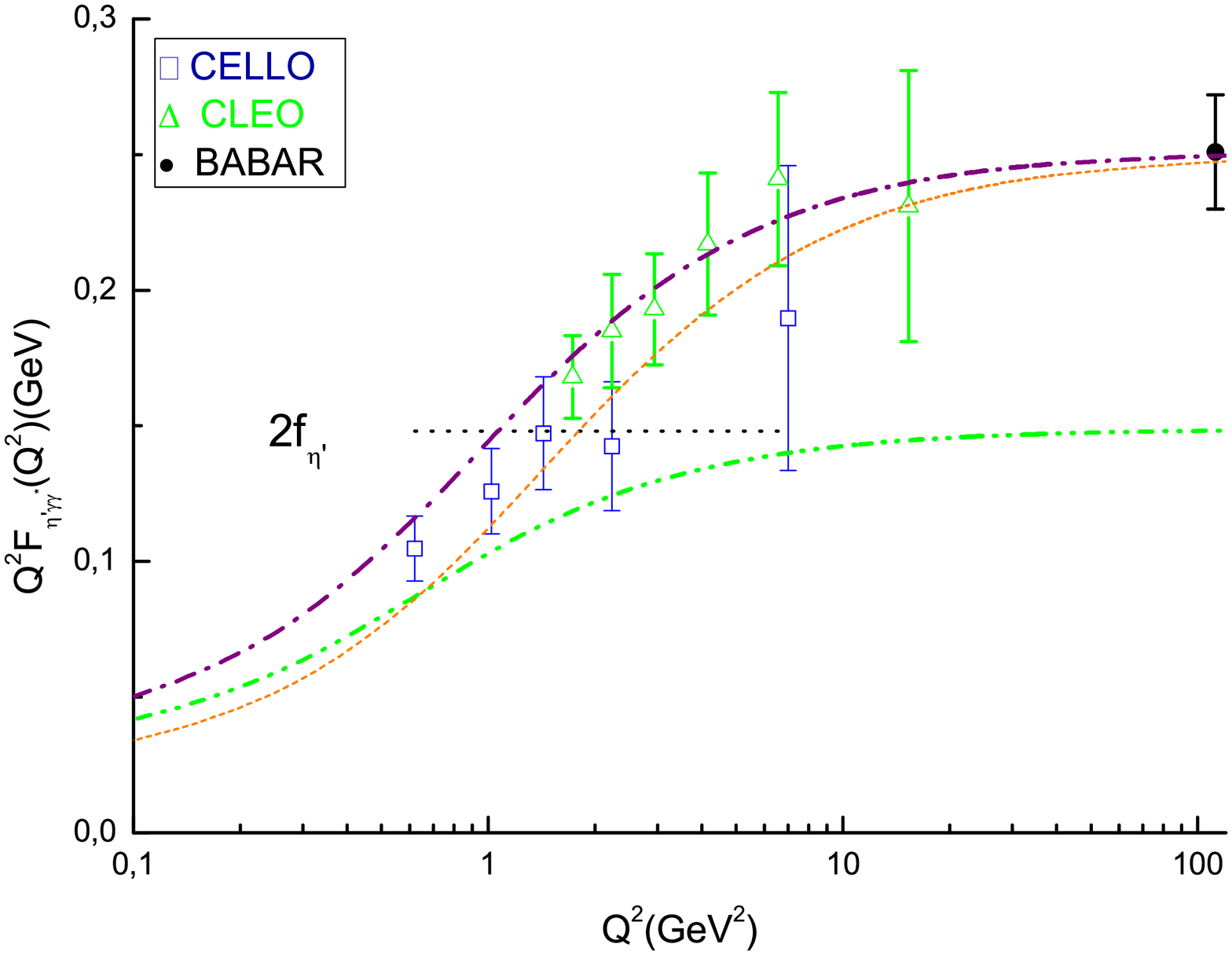}  \vspace*{-10mm}
\caption{{\protect\footnotesize The transition form factor $\protect\eta'
\rightarrow \protect\gamma^\ast\protect\gamma$. The data are from CELLO
\protect\cite{Behrend:1990sr}, CLEO \protect\cite{Gronberg:1997fj} and BABAR
\protect\cite{Aubert:2006cy} Collaborations. The dashed line is the perturbative QCD
asymptotic limit. The CLEO results obtained in different $\eta$ decay modes are averaged. (The notation for curves is explained in the text.)}}
\label{fig:eta1}
\end{figure}

At zero momentum transfer the transition form factors are fixed by
\begin{equation}
F_{P\gamma\gamma^*}^2(Q^2=0,0)=\frac{1}{(4\pi\alpha)^2}\frac{64\pi\Gamma(P\to\gamma\gamma)}{M_P^3},
\label{F0exp}\end{equation}
where $\alpha$ is the QED coupling constant, $M_P$ is the resonance mass and
$\Gamma(P\to\gamma\gamma)$ is the two-photon partial width of the meson $P$. The axial anomaly predicts
\begin{equation}
F_{P\gamma\gamma^*}(Q^2=0,0)\approx\frac{1}{4\pi^2f_P},
\label{F0the}\end{equation}
where $f_P$ is the meson decay constant. Under assumption of factorization the perturbative QCD provides the asymptotic behavior of the $F_{P\gamma\gamma^*}^2(Q^2,0)$ transition form factors as $Q^2\to\infty$ \cite{Brodsky:1981rp}
\begin{equation}
F_{P\gamma\gamma^*}(Q^2\to\infty,0)\sim\frac{2f_P}{Q^2}.
\label{Fas}\end{equation}
The perturbative QCD corrections to this expression at large momentum transfer are extremely small \cite{Chase:1979ck,Braaten:1982yp,Kadantseva:1985kb}.

To describe the soft nonperturbative region of $Q^2$ a simple interpolation between $Q^2\to0$ and $Q^2\to\infty$ limits has been proposed by Brodsky and Lepage (BL) \cite{Brodsky:1981rp}:
\begin{equation}
F^{\rm{BL}}_{\pi\gamma\gamma^*}(Q^2,0)=\frac{1}{4\pi^2f_P}\frac{1}{1+Q^2/(8\pi^2f^2_P)},
\label{Fbl}\end{equation}
where the values of $f_P$ are estimated from (\ref{F0exp}) and  (\ref{F0the}) \cite{Gronberg:1997fj}: $f_\pi=92.3$ MeV, $f_\eta=97.5$ MeV, $f_{\eta'}=74.4$ MeV.

The CLEO (and CELLO) collaboration parameterized their data by similar to (\ref{Fbl}) formula but with the pole mass being free fitting parameter \cite{Gronberg:1997fj}
\begin{equation}
F^{\rm{CLEO}}_{\pi\gamma\gamma^*}(Q^2,0)=\frac{1}{4\pi^2f_P}\frac{1}{1+Q^2/\Lambda_P^2},
\label{Fcleo}\end{equation}
where $\Lambda_\pi=776\pm22$ MeV, $\Lambda_\eta=774\pm29$ MeV, and $\Lambda_{\eta'}=859\pm28$ MeV.

The asymptotics (\ref{Fas}) are shown by dotted lines, the BL interpolations are given by dot-dot-dashed lines, and the CLEO parametrization extrapolated to higher momentum transfer is shown by dot-dashed lines  in Figs. 1-3. We see that the QCD inspired expression (\ref{Fbl}) works well only for the $\eta$ meson form factor (Fig. 2), while the CLEO parametrization (\ref{Fcleo}) underestimates the large $Q^2$ behavior. From other side the CLEO parametrization well describes the $\eta'$ meson form factor (Fig. 3), but the BL expression strongly underestimates the large $Q^2$ behavior. We still have good description of the $\eta'$ meson form factor by BL formula if one takes $f_{\eta'}=125$ MeV (short dashed line in Fig. 3), but then the normalization is incorrect.

For the $\eta$ and $\eta'$ mesons the parametrizations (\ref{Fbl}) and (\ref{Fcleo}) reflect correctly the experimental data at large $Q^2$ at qualitative level. This is not the case for the pion form factor showing the growth at large  $Q^2$ unexpected from the QCD factorization approach (Fig. 1).
However, this growth is easy to explain \cite{Dorokhov:2009dg}  in the context of the quark model \cite{Quark2}. Within this model, the pion form factor is given by the quark-loop (triangle)
diagram with momentum independent constituent quark mass serving as an infrared regulator. The form factor has double logarithmic asymptotics at large $Q^2$: $log^2(Q^2/M_q^2)$ and is given by \cite{Quark2}
\begin{eqnarray}
&&F_{\pi\gamma\gamma^*}(Q^2,0)=\frac{1}{4\pi^2f_\pi}\frac{m_\pi^2}{m_\pi^2+Q^2}\frac{1}{%
2\arcsin^2(\frac{m_\pi}{2M_Q})}  \nonumber \\
&&\cdot\{2\arcsin^2(\frac{m_\pi}{2M_Q})+\frac{1}{2}\ln^2\frac{\beta_Q+1}{%
\beta_Q-1}\}.  \label{Ftt}
\end{eqnarray}
where $\beta_Q=\sqrt{1+\frac{4M_Q^2}{Q^2}}$.

Within perturbative QCD the possible scenario to explain the form factor
growth faster than expected is to assume that the pion distribution
amplitude is flat and there is no QCD evolution \cite{Radyushkin:2009zg}. In
the proposed model the pion transition form factor is
\begin{eqnarray}
&&F_{\pi\gamma\gamma^*}(Q^2,0)=\frac{2}{3}\frac{f_\pi}{Q^2}  \nonumber \\
&&\cdot\int_0^1\frac{dx}{x}\left[1-\exp{\left(-\frac{xQ^2}{2\sigma (1-x)}%
\right)}\right].  \label{Frad}
\end{eqnarray}
and has logarithmically enhanced asymptotic behavior $\sim \log{%
\left(1+Q^2/\sigma\right)}$.

The solid and dashed lines in Fig. 1 are the pion transition form factor calculated from Eq. (\protect\ref{Ftt}) with $M_Q=135$ MeV and Eq. (\ref{Frad}) with $\sigma = 0.48$ GeV$^2$, respectively. They practically coincide and describes well the BABAR data. Note that these logarithmically enhanced models are not able to describe $\eta(')$ form factors.

The possible origin in difference of asymptotic behavior of the pion and $\eta(')$ meson form factors is with flavor composition of mesons\footnote{I am indebted to N.I. Kochelev for discussion this point.}. The pion consists of almost massless $u,d$ quarks, while the $\eta(')$ mesons include also $s$ quark. The $s$ quark with mass $m_s$ of order $\Lambda_{QCD}$ may be considered as a heavy one. Recently the similar behavior like predicted by (\ref{Fcleo}) was found for $\gamma\gamma^*\to\eta_c$ transition form factor measured by the BABAR collaboration for the range $Q^2=2-50$ GeV$^2$ \cite{Druzhinin:2009gq}. The corresponding fitted mass parameter is $\Lambda_{\eta_c}=2.92\pm16$ GeV.

It is important to confirm the theoretical base for maximally model independent prediction of the branchings (see Table) by getting more precise data on the pion
transition form factor in asymmetric as well in symmetric kinematics in wider region of momentum transfer that is soon expected from the BABAR, BELLE (at large momentum transfer) and KEDR (at small momentum transfer) collaborations.

We are grateful to the Organizers and personally S. Dubnicka, A.-Z. Dubnickova, E. Bartos, M. Hnatic for a nice meeting and kind invitation to present our results. The author thanks S.B. Gerasimov, V.P. Druzhinin, N.I. Kochelev, S.V. Mikhailov, A.V. Radyushkin for discussions on the interpretation of the high momentum transfer data for the pseudoscalar meson transition form factors.

\end{document}